\documentclass[graybox]{svmult}

\usepackage{type1cm}        % activate if the above 3 fonts are
\usepackage{makeidx}         % allows index generation
\usepackage{graphicx}        % standard LaTeX graphics tool
\usepackage{multicol}        % used for the two-column index
\usepackage[bottom]{footmisc}% places footnotes at page bottom

\usepackage{newtxtext}       % 
\usepackage{newtxmath}       % selects Times Roman as basic font

\usepackage{listings}
\usepackage[frozencache,cachedir=.]{minted}

\usepackage{url}
\usepackage{xspace}
\usepackage{graphics}
\usepackage{bbold}
\usepackage{caption}
\usepackage{colortbl}
\usepackage{multirow}
\usepackage{fancyvrb,keyval,ifthen}
\usepackage{booktabs}
\usepackage{enumitem}
\setlist{nolistsep}
\usepackage{array}
\usepackage{algorithm}
\usepackage{algorithmic}
\usepackage{pifont}
\usepackage{comment}

\newcommand{\markyes}{\ding{51}}%
\newcommand{\markno}{\ding{55}}%

\makeindex             % used for the subject index
\begin{document}

\title*{Flexible Multi-Dimensional FFTs for Plane Wave Density Functional Theory Codes}
\author{Doru Thom Popovici, Mauro Del Ben, Osni Marques, Andrew Canning}
\institute{Doru Thom Popovici, Mauro Del Ben, Osni Marques, Andrew Canning \at Lawrence Berkeley National Lab, 1 Cyclotron Rd, USA, 94720, \email{dtpopovici@lbl.gov, mdelben@lbl.gov, oamarques@lbl.gov, acanning@lbl.gov}}

\maketitle
\vspace{-25mm}
\abstract{Multi-dimensional Fourier transforms are key mathematical building blocks that appear in a wide range of applications from materials science, physics, chemistry and even machine learning. Over the past years, a multitude of software packages targeting distributed multi-dimensional Fourier transforms have been developed. Most variants attempt to offer efficient implementations for single transforms applied on data mapped onto rectangular grids. However, not all scientific applications conform to this pattern, i.e. plane wave Density Functional Theory codes require multi-dimensional Fourier transforms applied on data represented as batches of spheres. Typically, the implementations for this use case are hand-coded and tailored for the requirements of each application. In this work, we present the Fastest Fourier Transform from Berkeley (FFTB) a distributed framework that offers flexible implementations for both regular/non-regular data grids and batched/non-batched transforms. We provide a flexible implementations with a user-friendly API that captures most of the use cases. Furthermore, we provide implementations for both CPU and GPU platforms, showing that our approach offers improved execution time and scalability on the HP Cray EX supercomputer. In addition, we outline the need for flexible implementations for different use cases of the software package.}

\section{Introduction}
The discrete Fourier transform has proven to be an important mathematical kernel that is widely used in a multitude of applications. For example, applications from molecular dynamics~\cite{Plimpton97particle-meshewald, plimpton2012computational, plimpton2017ffts}, material sciences~\cite{lebensohn2012elasto, lebensohn2001n, lee2011modeling}, ~\cite{vay2018warp, vay2013domain}, quantum mechanics~\cite{kendall2000high,valiev2010nwchem,straatsma2011advances, canning2000parallel} and even machine learning~\cite{li2020fourier, wen2022u} require flexible and efficient implementations of multi-dimensional Fourier transforms that can be executed on parallel and distributed systems. In recent years, a multitude of software packages such as FFTE~\cite{DBLP:conf/ppam/Takahashi09}, heFFTe~\cite{ayala2020heffte}, elemental FFT~\cite{popovici2020flexible}, FFTU~\cite{koopman2023minimizing} have attempted to offer efficient implementations targeting multi-node CPU and/or GPU systems. Table~\ref{tab:dags} outlines the different software packages and some of their characteristics. Most frameworks focus on single three-dimensional Fourier transforms applied on data mapped to cuboid grids. However, some scientific applications such as plane wave Density Functional Theory calculations do not fit within this mold. Density Functional Theory codes require batched three-dimensional transforms applied on non-cuboid shaped data, hence demanding a specialized implementations.

\begin{table}[t]
    \centering
    \begin{tabular}{c c c c c c}
    \toprule
     Software &  Platform & Transform Type & Input/Output Type & Processing Grid & Batching  \\ 
    \toprule 
    FFTE~\cite{FFTE} & CPU & CtoC/RtoC & Cuboid & 1D/2D & \markno \\
    heFFTe~\cite{ayala2020heffte} & CPU/GPU & CtoC/RtoC & Cuboid & 1D/2D & \markno \\
    FFTX~\cite{franchetti2018fftx} & CPU/GPU & CtoC/RtoC & Cuboid & 1D/2D & \markno \\
    FFTU~\cite{koopman2023minimizing} & CPU & CtoC & Cuboid & 1D/2D & \markno \\
    elemental FFT~\cite{popovici2020flexible} & CPU & CtoC & Cuboid & 1D/2D/3D & \markno \\
    FFTB (ours) & CPU/GPU & CtoC & Cuboid/Sphere & 1D/2D/3D & \markyes\\
    \bottomrule
    \end{tabular}
    \caption{The table contains different implementations for distributed Fourier transforms and some of their characteristics. We outline four characteristics: the transform type (complex to complex (CtoC) or real to complex (RtoC), the input/output type (cuboid or sphere data), the shape of the processing grid (1D, 2D, or 3D), the batching mechanism (present or not).}
    \label{tab:dags}
    \vspace{-3mm}
\end{table}

First-principles Density Functional Theory calculations in the Kohn-Sham formalism are widely used to study the electronic structure of materials. The Kohn-Sham single particle equation is expressed as
\begin{align}
    \hat{H}\psi_i(r) = \epsilon_i\psi_i(r),
    \label{eq:ks}
\end{align}
where $\psi_i(r)$ is the $i$-th electron orbital (wavefunction) in the system, $\epsilon_i$ represents the electron orbital's eigen energy and $\hat{H}$ is the Hamiltonian of the system describing the different interactions. Each wavefunction spans the entire 3D real space $r$. However, solving Equation~\ref{eq:ks} requires the wavefunctions either to be discretized on a real space grid or expressed in some basis set. One of the most widely used basis sets is the plane wave basis set~\cite{PhysRevLett.55.2471}, which discretizes the wavefunctions using a Fourier series expansion. It is conventional to truncate the plane wave basis set within an energy cutoff and thus only frequency coefficients within a cut-off sphere are preserved for computation. Given that some operations applied on the wavefunctions are cheaper in real space, inverse and forward Fourier transforms are required to change from frequency to real space and back. The three-dimensional Fourier transforms need extra steps to deal with the non-cuboid shape of the data. In addition, the transform should be applied on batches of spheres which are packed together to reduce latency issues in the communications. 

Most plane wave Density Functional Theory codes~\cite{giannozzi2017advanced, gygi2008architecture, gonze2009abinit, andrade2021inq} either implement their own Fourier transform variant or resort to padding the spheres to cuboid grids and use off-the-shelf frameworks. The first approach provides efficient implementations, at the cost of having a rigid solution that is specifically tailored to each application. The second approach trades performance for flexibility, requiring redundant computation and communications. Figure~\ref{fig:strong_scaling} outlines the differences in execution performance measured in terms of strong scaling behavior between the two approaches, both implemented within our proposed framework. As expected, the implementation that does not pad the data provides better execution time, while still scaling to many compute nodes. In this work, we present a flexible implementation that allows users to express different input shapes and then efficiently applies the computation and data movement, targeting both CPUs and GPUs. Furthermore, we analyze key aspects of our approach and then show scaling results for our implementation given certain use cases from Density Functional Theory codes.

{\bf{Contributions.}} This work makes the following contributions:
\begin{enumerate}
    \item We present FFTB (the Fastest Fourier Transform from Berkeley) a distributed multi-dimensional Fourier transform for classical and plane wave Density Functional Theory codes.
    \item We present a flexible implementation that allows different shapes of the data to be easily expressed by users.
    \item We present an efficient implementation capable of  excellent parallel scalability on both CPU and GPU distributed systems.
\end{enumerate}

\section{Background}
In this section, we briefly present the Fourier transform, focusing on the three-dimensional implementation. We give a brief introduction to the classical three-dimensional transform applied on data stored on cuboid grids, and then we discuss the plane wave Fourier transform applied on batches of spheres.

\subsection{The Fourier Transform}

The discrete one-dimensional Fourier transform (we use the DFT abbreviation for the discrete Fourier transform) is a matrix-vector multiplication, where given the input $x$ of size $n$, the output $y$ is obtained as
\begin{align}\label{eq:1ddft}
    y = DFT_n\cdot x.
\end{align}
The $DFT_n$ is the $n \times n$ Fourier dense matrix, defined as 
\begin{align}
    DFT_n = \begin{bmatrix}\omega^{l k}_n\end{bmatrix}_{0\leq l, k < n},
\end{align}
where $\omega_n = e^{-j\frac{2\pi}{n}}$ is the complex root of unity. The computation of the DFT is implemented using the Fast Fourier Transform (DFT), where instead of performing $O(n^2)$ complex arithmetic operations by doing the matrix-vector multiplication, a recursive decomposition of the DFT matrix is performed to obtain an $O(n \log(n))$ algorithm. The most widely known of these algorithms is the Cooley-Tukey algorithm~\cite{cooley1965algorithm}, which can be described concisely as a factorization of the $DFT_n$ matrix for $n = n_0\times n_1$. For more details on the decomposition, notation and implementations, we recommend the reader follow the works~\cite{popovici2018approach, popovici2020flexible, popovici2021systematic}.

\begin{figure}[t]
    \centering
    \includegraphics[width=0.6\textwidth]{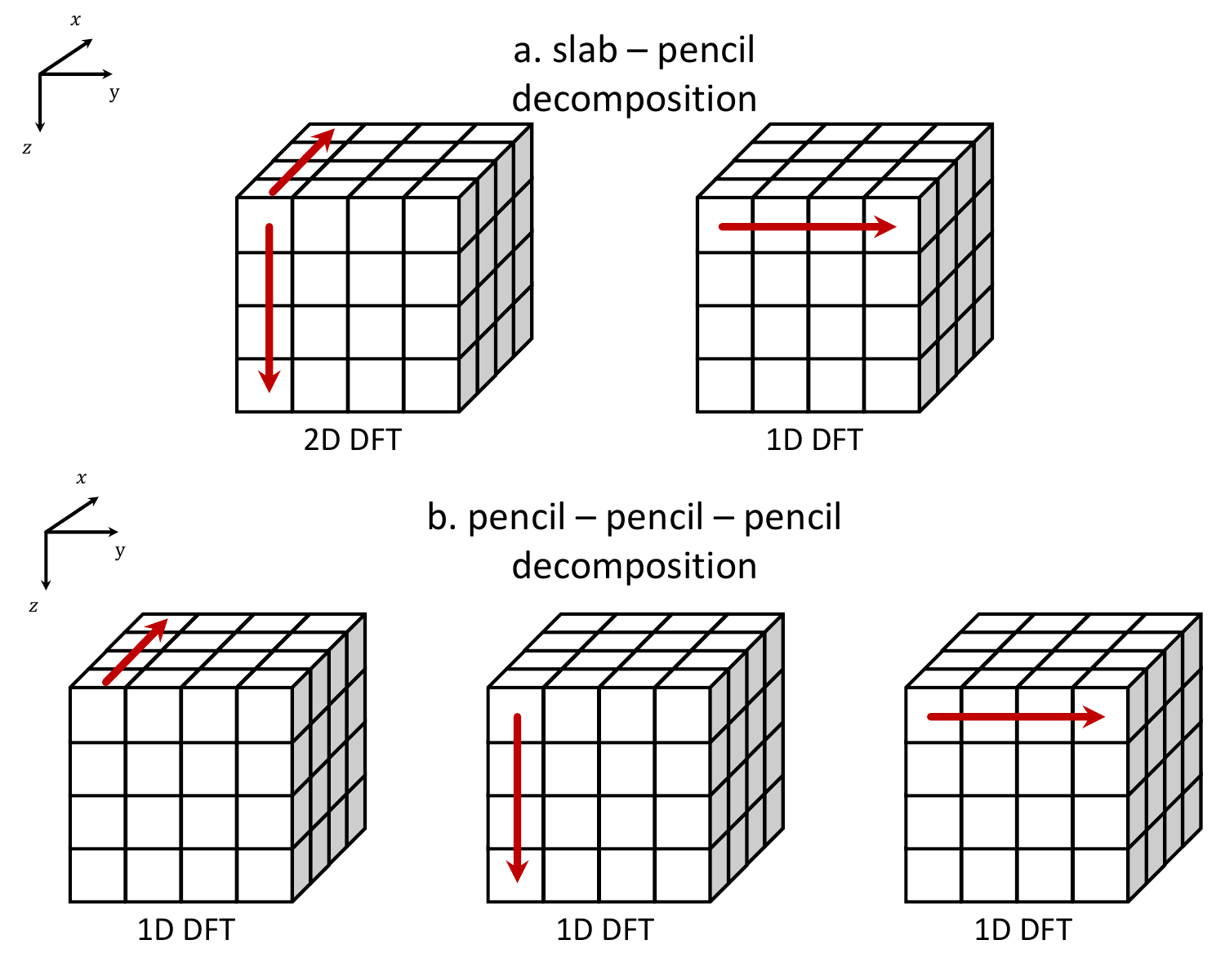}
    \caption{Two algorithms used to compute a three dimensional Fourier transform. The top algorithm decomposes the computation as a batch of 2D transforms applied in the $xy$-plane and a batch of 1D transforms applied in the $z$ dimension. The bottom algorithm, applies the three-dimensional Fourier transform as three groups of 1D Fourier transforms applied in each dimension of the input three dimensional tensors.}
    \label{fig:fft}
    \vspace{-2mm}
\end{figure}

Multi-dimensional Fourier transform can also be expressed as a matrix-vector multiplication. A three-dimensional DFT of size $n_0\times n_1\times n_2$ is expressed as
\begin{align}
    y = DFT_{n_0\times n_1\times n_2}\cdot x,
    \label{eq:3dfft}
\end{align}
where the $DFT_{n_0\times n_1\times n_2}$ represents the complex DFT matrix. Any multi-dimensional DFT can be expressed in terms of multiple 1D DFTs and/or multi-dimensional DFTs of lower dimensions, which themselves are expressed as 1D DFTs. For example, Figure~\ref{fig:fft} shows (a) the slab-pencil algorithm that decomposes the 3D DFT into a batch of 2D DFTs followed by a batch of multiple 1D DFTs, and (b) the pencil-pencil-pencil algorithm that splits the 3D DFT into three batches of 1D DFTs, where each 1D DFT is applied in the corresponding three dimensions. The matrix-vector notation can be extended to one dimensional Fourier transforms applied on higher dimensional tensors. 

The slab-pencil algorithm sees the input (output) column vectors $x$ ($y$) as two dimensional matrices $\tilde{x}$ ($\tilde{y}$) of size $\left(n_0\times n_1\right)\times n_2$. The Fourier computation can mathematically be expressed as
\begin{align}
    \tilde{y} = \left(DFT_{n_0\times n_1}\cdot\tilde{x}\right)\cdot DFT_{n_2}
\end{align}
Data is stored in column major, hence the 2D DFT is applied on the columns, while the 1D DFT is applied on the rows. Subsequently, the 2D DFT applied can further be decomposed as 1D DFTs. Each column from matrix $\tilde{x}$ can be viewed as a two-dimensional matrix of size $n_0\times n_1$.

The pencil-pencil-pencil algorithm views the input (output) vectors as 3D cubes $\hat{x}$ ($\hat{y}$) of size $n_0\times n_1\times n_2$. The input (output) column vector $x$ ($y$) is decomposed into $n_2$ groups of $n_1$ subgroups of size $n_0$. Hence, the three-dimensional cube $\hat{x}$ is viewed as a matrix of matrices such that
\begin{align}
    \hat{x} = \begin{bmatrix}\tilde{x}_0| & \tilde{x}_1| & \ldots| & \tilde{x}_{n_2-1}\end{bmatrix},
\end{align}
where each $\tilde{x}_i$ is the 2D matrix of size $n_0\times n_1$ for all values $0\leq i < n_2$. Mathematically, the pencil-pencil-pencil algorithm is expressed as
\begin{align}
    \tilde{t}_i &= \left(DFT_{n_0}\cdot\tilde{x}_i\right)\cdot DFT_{n_1},\forall 0\leq i < n_2\\\nonumber
    \hat{y} &= \begin{bmatrix}\tilde{t}_0 & | & \tilde{t}_1 & | & \ldots &|& \tilde{t}_{n_2 - 1}\end{bmatrix}\cdot DFT_{n_2}
\end{align}
where the $DFT_{n_0}$ is applied in the depth dimension, the $DFT_{n_1}$ is applied in the column dimension and finally the $DFT_{n_2}$ is applied in the row dimension. Data is stored in column major order and therefore the dimension corresponding to $n_0$ is laid out in the fastest dimension in memory, while the dimension corresponding to $n_2$ is laid out in the slowest dimension in memory. The order in which the 1D and/or 2D DFTs are applied can be permuted for both the slab-pencil and pencil-pencil algorithms.

\begin{figure}[t]
    \centering
    \includegraphics[width=0.5\textwidth]{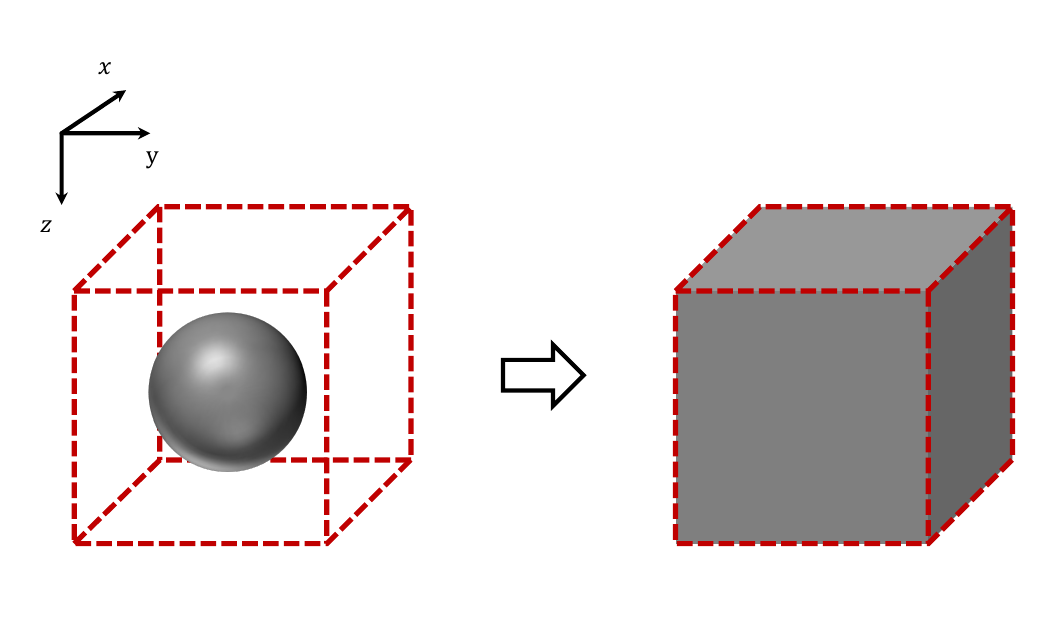}
    \caption{Each wavefunction is decomposed using a Fourier series expansion and only the complex coefficients within a cut-off sphere are kept. The 3D Fourier computation requires the data to be on a cuboid grid. As such, the data can be padded with zeros to a cube typically of width twice the diameter of the sphere.}
    \label{fig:planewave_fft_full}
    \vspace{-2mm}
\end{figure}

\subsection{The Plane Wave Fourier Transform}

\begin{figure}[t]
    \centering
    \includegraphics[width=0.9\textwidth]{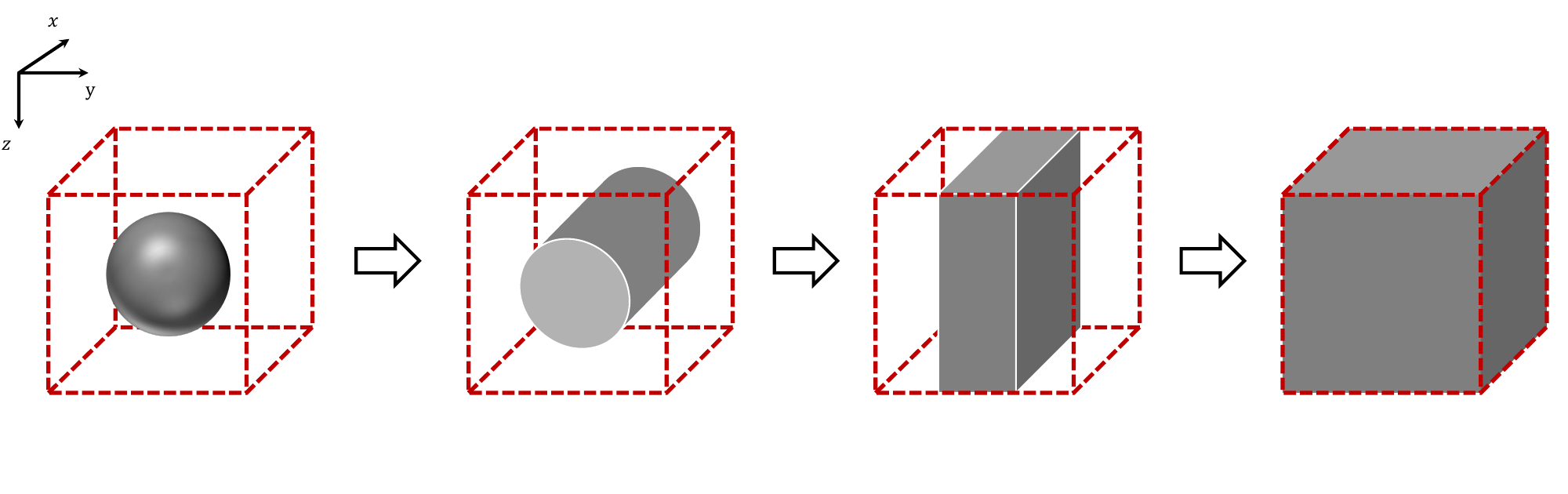}
    \caption{The padding operation can also be split by dimensions. The padding is done in the $x$-dimension first, followed by the $y$-dimension and $z$-dimension. The 3D Fourier transform has a similar decomposition, so after each padding operation, the 1D transform can be immediately applied. Exploiting the structure of the data will reduce the amount of data that is being communicated and computed upon.}
    \label{fig:planewave_fft_step_by_step}
    \vspace{-2mm}
\end{figure}

Solving Equation~\ref{eq:ks} requires the wavefunctions to be discretized on a real space grid or expressed in some basis set. One of the most widely used basis sets is the plane wave basis set~\cite{PhysRevLett.55.2471}. 
For the simplest case, 
%of a gamma point calculation,
the discretization of the $\psi_i(r)$ can be expressed as a Fourier series expansion such that
\begin{align}
    \psi_i(r) = \sum_g c_i(g) e^{j g r},
\end{align}
where $g$ represents the reciprocal lattice vectors of the supercell to be studied and $j=\sqrt{-1}$. The $c_i(g)$ terms represent the plane wave complex coefficients. It is conventional to truncate the plane wave basis set within an energy cutoff $E_{cut}$ such that
\begin{align}
    \frac{1}{2}|g|^2 \leq E_{cut}.
\end{align}
Therefore, only the $g$ vectors within a cut-off sphere are used in the above expansion as shown in Figure~\ref{fig:planewave_fft_full}. The corresponding complex coefficients are stored as a column vector $\psi_i$. In addition, an offset array is typically constructed to specify the correct location of the data in the sphere. The same offset array can be used for all wavefunctions $\psi_i$. Equation~\ref{eq:ks} can be solved for all $\psi_i$ wavefunctions using a Conjugate Gradient (CG) algorithm (there are other algorithms that have been used to determine the solutions). The same CG steps are applied for all the wavefunctions $\psi_i$. Therefore, the wavefunctions can be batched together such that
\begin{align}
    \Psi = \begin{bmatrix}\psi_0 & | & \psi_1 & |& \ldots&|& \psi_{N_b-1}\end{bmatrix},
\end{align}
where $N_b$ represents the number of wavefunctions or the number of bands. The all-band algorithm converts most of the computation steps in the CG algorithm from matrix-vector multiplication to matrix-matrix multiplications. Furthermore, the Fourier transforms are applied on batches of data and not on a single column vector at a time.

The Fourier transform requires data to be placed on a cuboid grid. As such, all the wavefunction spheres must be zero-padded. Figure~\ref{fig:planewave_fft_full} shows the case that zero-pads the entire sphere by embedding it into a cube of width twice the size of the diameter of the sphere (this is a requirement from the solvers). The 3D Fourier computation is applied on the entire cube. While the advantage of this approach is that off-the-shelf libraries can be used to compute the 3D Fourier transform, the disadvantage is that the amount of data is increased by almost $16$ times, affecting computation and data movement. Alternatively, the zero-padding can be done in stages as outlined in Figure~\ref{fig:planewave_fft_step_by_step}, where each dimension is gradually padded with zeros. Splitting the 3D Fourier transform into the corresponding 1D transforms, enables us to fuse the 1D transforms with the padding operations. The advantage of this approach is that it keeps the amount of redundant work to a minimum. However, none of the classical implementations can provide support for this implementation.

\subsection{Connections to Multi-Linear Algebra}

In the work by Popovici et al.~\cite{popovici2021systematic}, it has been shown that Fourier transforms as well as the padding operations can be expressed as linear algebra operations. As such, utilizing a unifying notation for multiple computation steps, enabled the possibility of fusing the padding operations, the Fourier transforms and the linear algebra operations. This in turn provided a boost in performance on shared memory systems. In the work by Popovici et al.~\cite{popovici2020flexible}, a flexible implementation for distributed 3D Fourier transforms has been presented. The approach expressed the 3D Fourier transforms as multi-linear algebra operations and borrowed techniques from tensor community to provide a distributed Fourier transform using 3D volumetric decomposition on large number of CPUs. Finally, the work by Franchetti el al~\cite{franchetti2018fftx} has provided an API for Fourier-based operations seen in applications like Density Functional Theory calculations, plasma beam simulations~\cite{vay2018warp} or methods of local corrections~\cite{mccorquodale2007local}, with the goal of enabling cross domain optimizations. We build on these approaches and develop our own distributed Fourier framework that targets both cuboid and non-cuboid data shapes.

\section{A Flexible Framework for Multi-Dimensional Fourier Transforms}
\begin{figure}[t]
    \centering
    \includegraphics[width=0.7\textwidth]{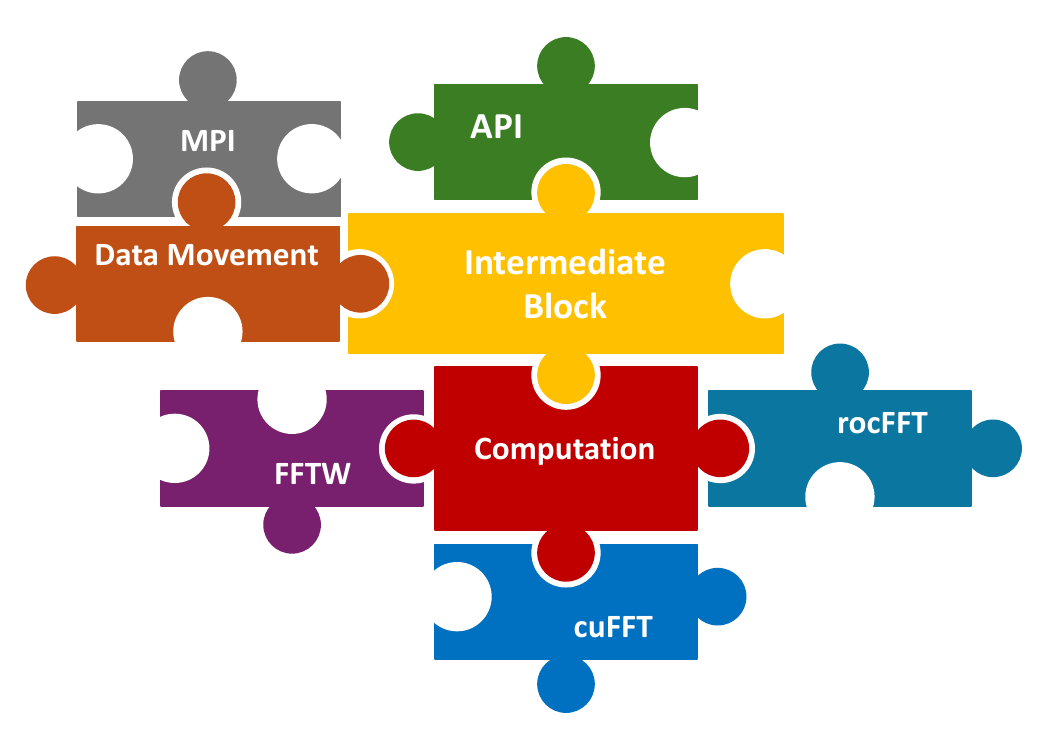}
    \caption{The structure of FFTB. The API (green block) contains the main functionalities for describing distributed Fourier transforms. The intermediate block (yellow block) creates and links the stages of the Fourier transform based on the distribution of the inputs/outputs. The main stages are either local computation stages (red block) or data movement stages (orange block).}
    \label{fig:overview}
    \vspace{-2mm}
\end{figure}

In this section, we describe our framework entitled the Fastest Fourier Transform from Berkeley (FFTB). We provide an overview of our framework outlining the key components. Then we present the capabilities of our approach by providing some simple examples for distributed 3D Fourier transforms.

\subsection{Overview}

We develop a modular implementation for the FFTB framework. In the following paragraphs, we will describe each component, outlining the key features of each module. We show how the framework can be extended to different platforms.

{\bf{Computational Layout Descriptor.}} We provide a C/C++ API to describe distributed multi-dimensional Fourier transforms. The API provides the necessary functionalities to specify the processing grid, the input and output tensors and how they are distributed across the processing grid. Finally, the API specifies the necessary constructs to create, execute and clean-up multi-dimensional Fourier transforms. In the following sub-sections, we provide two examples. The first example focuses on the classical 3D Fourier transform for data stored on cuboid grids. The second example shows the extensions needed to express batched Fourier transforms specific for the plane wave Density Functional Theory codes.

{\bf{Distributed Fourier Transform Creation.}} The description of the distributed Fourier transform is translated to an intermediate representation. The intermediate block (yellow block) analyses the distribution patterns of the input/output tensors and constructs the necessary compute and communicate stages. Basically, this part stitches together the end-to-end implementation of the required Fourier transform. The current FFTB implementation accepts some predefined patterns that are used in most scientific applications. The framework will raise an exception if the provided patterns are not within the predefined list. We leave as future work, the approach of deciding the stages based on the distribution of the input/output tensors.

{\bf{Local Computation and Data Movement Specification.}} The local computation is represented by the 1D or 2D Fourier transforms that are applied on the data. We abstract the computation and create a couple of functions to specify in which dimensions the 1D or 2D transforms are applied in. The abstractions are replaced with actual function calls from off-the-shelf libraries like FFTW, cuFFT and rocFFT. Similarly, data movement is also abstracted away (typically, Fourier transforms required \texttt{alltoall} MPI collectives). We provide compilation flags and let users specify what libraries they want their libraries to be linked against.

In the following sub-sections, we provide two examples on how to use the API. We provide two examples for describing distributed Fourier transforms. We start with the 3D Fourier transform applied on cuboid grids and continue with a batched 3D transform on non-cuboid data (wavefunction spheres).

\subsection{An API to Express Fourier Transforms on Processing Grids}

\begin{figure}[t]
    \centering
    \includegraphics[width=0.5\textwidth]{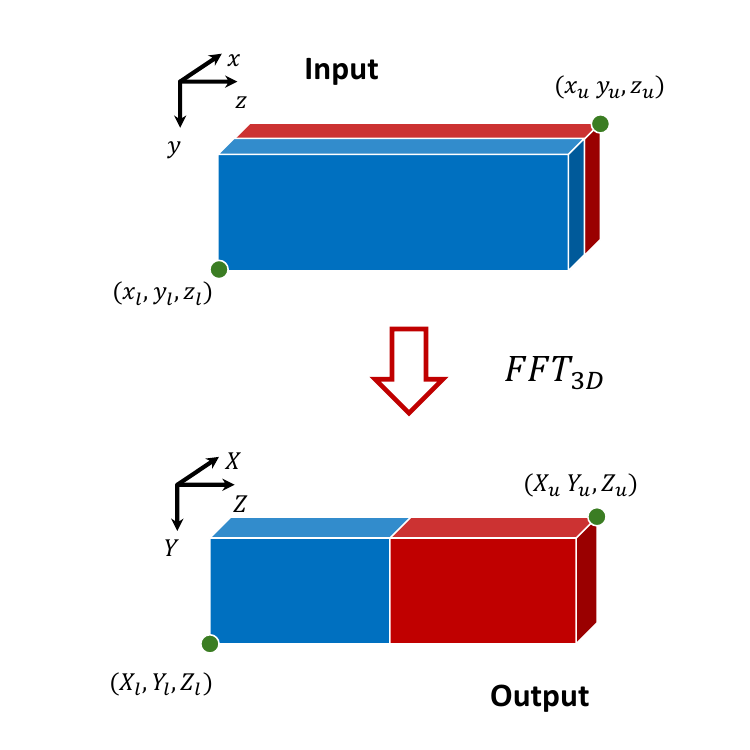}
    \caption{A 3D Fourier transform applied on an input tensor distributed in the $x$-dimension. The result of the Fourier transform is a tensor that is distributed in the $z$-dimension.}
    \label{fig:fft_distrib}
    \vspace{-2mm}
\end{figure}

Figure~\ref{fig:fft_distrib} depicts a 3D Fourier transform applied on three-dimensional tensors distributed across a given processing grid. The input/output tensors are defined by specifying the bound domains, like the approach presented by the FFTX project~\cite{franchetti2018fftx}. Each domain is defined by specifying the points corresponding to opposite corners of cuboid volume, i.e. $(x_l, y_l, z_l)$ and $(x_u, y_u, z_u)$ for the input tensor and $(X_l, X_l, X_l)$ and $(X_u, X_u, X_u)$ for the output tensor. The tensors are created by specifying the domains and the distribution across the processing grid. The input tensor is distributed in the $x$-dimension, while the output tensor is distributed in the $z$-dimension. The grid, the tensors and their distribution are sufficient to create the corresponding distributed 3D Fourier transform.

\begin{figure}[t]
\centering
\begin{minted}[mathescape,
               escapeinside=||,
               linenos,
               numbersep=5pt,
               frame=lines,
               framesep=2mm]
{c++}
// create processing grid
std::vector<int> procs{16};
grid g = grid(procs, MPI_COMM_WORLD);

// create input tensor
std::vector<int> point_in_lower{0, 0, 0};
std::vector<int> point_in_upper{255, 255, 255};

std::vector<domain> dom_in;
dom_in.push_back(domain(point_in_lower, point_in_upper));
tensor ti = tensor(dom_in, "x{0} y z", g);

// create output tensor
std::vector<int> point_out_lower{0, 0, 0};
std::vector<int> point_out_upper{255, 255, 255};

std::vector<domain> dom_out;
dom_out.push_back(domain(point_out_lower, point_out_upper));
tensor to = tensor(dom_out, "X Y Z{0}", g);

// create fft operation
std::vector<int> size{256, 256, 256};
fftb fx = fftb(sizes, to, "X Y Z", ti, "x y z", g);
\end{minted}
\caption{Code snippet used to describe a distributed 3D Fourier transform of size $256^3$. The first lines describe the creation of the processing grid, the next lines specify the input and output tensors and how they are distributed, the last lines specify the creation of a 3D Fourier transform.}
\label{alg:fft_classic}
\vspace{-5mm}
\end{figure}

Figure~\ref{alg:fft_classic} describes the step by step process for creating the processing grid, specifying the input and output tensors, and creating the Fourier transform. First, lines $2$-$3$ specify the creation of the processing grid. We create a 1D grid containing $16$ processors. Second, lines $6$-$11$ specify the creation and distribution of the input tensor \texttt{ti}. We create the lower and upper points of the bounding domain (lines $6$-$7$). We then create the domain on lines $9$-$10$. Finally, we create the distributed tensor. The tensor takes the domain, a string specifying the dimensions and the distribution and the processing grid \texttt{g}. In the example, we specify that the tensor \texttt{ti} is a three-dimensional tensor with dimensions $x$, $y$, and $z$ and only the $x$ dimension is distributed across the $0$-th dimension of the processing grid. We use the elemental cyclic distribution described in~\cite{popovici2020flexible} to distribute the data (data in each dimension is distributed in a round robin fashion at the granularity of one element). The current implementation does not support the blocked distribution.

Thirdly, we describe the output tensor \texttt{to} in a similar manner (lines $14$-$19$). While the input tensor \texttt{ti} is distributed in the $x$-dimension, the output tensor \texttt{to} is distributed in the $z$-dimension. Finally, lines $22$-$23$ outline the code snippet that creates the distributed Fourier transform. We specify the size, the input and output tensors and the grid. From this specification, the framework decides on the most suited implementation. Once the Fourier transform is constructed, the \texttt{fx} object can be used to execute the computation. 

\subsection{Extensions for the Plane Wave Fourier Transform}

\begin{figure}[t]
    \centering
    \includegraphics[width=0.6\textwidth]{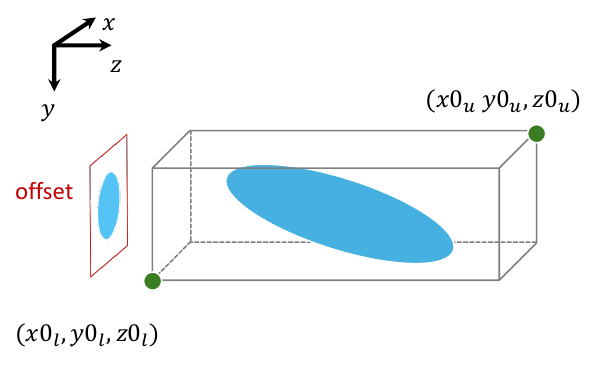}
    \caption{Each wavefunction is discretized using a Fourier series expansion. Some of the coefficients are discarded based on the cut-off energy $E_{cut}$. The sphere is bounded within a bounding domain. An offset array is constructed by projecting the data points on the $xy$-plane. That offset array is used to access the actual data points.}
    \label{fig:spheres}
    \vspace{-2mm}
\end{figure}

The wavefunctions used in Density Functional Theory calculations are discretized using a Fourier series expansion. Typically, not all Fourier coefficients are saved. Those coefficients for which the frequency is outside a cut-off energy are discarded. As such, the plane wave representation of the wavefunctions will be as that of batches of spheres as the one outlined in Figure~\ref{fig:spheres} for a single sphere. The sphere can be bounded by a domain. Furthermore, projecting the data on the $xy$-plane an offset array can be constructed, which resembles the Compressed Sparse Row format. Software like Quantum Espresso~\cite{giannozzi2017advanced} use such a structure to store the wavefunctions. We can use this information to extend the functionality of our API. This in turn will allow users to specify offset arrays within bounding domains.

\begin{figure}[t]
\centering
\begin{minted}[mathescape,
               escapeinside=||,
               linenos,
               numbersep=5pt,
               frame=lines,
               framesep=2mm]
{c++}
// initialize offset array
offsets = ...

// create processing grid
std::vector<int> procs{16};
grid g = grid(procs, MPI_COMM_WORLD);

// create batch dimension
std::vector<int> point_b_lower{0};
std::vector<int> point_b_upper{128};

// create input tensor
std::vector<int> point_in_lower{0, 0, 0};
std::vector<int> point_in_upper{255, 255, 255};

std::vector<domain> dom_in;
dom_in.push_back(domain(point_b_lower, point_b_upper));
dom_in.push_back(domain(point_in_lower, point_in_upper, offsets));
tensor ti = tensor(dom_in, "b x{0} y z", g);

// create output tensor
std::vector<int> point_out_lower{0, 0, 0};
std::vector<int> point_out_upper{255, 255, 255};

std::vector<domain> dom_out;
dom_out.push_back(domain(point_b_lower, point_b_upper));
dom_out.push_back(domain(point_out_lower, point_out_upper));
tensor to = tensor(dom_out, "B X Y Z{0}", g);

// create fft operation
std::vector<int> size{256, 256, 256};
fftb fx = fftb(sizes, to, "X Y Z", ti, "x y z", g);
\end{minted}
\caption{Code snippet used to describe a distributed planewave Fourier transform of size $256^3$. The declaration is similar to the classical implementation. The main differences are specification of the batch dimension, and the addition of an offset array to the domain of the input tensor.}
\label{alg:fft_spheres}
\vspace{-5mm}
\end{figure}

Figure~\ref{alg:fft_spheres} describes the step by step process for creating the processing grid, specifying the input and output tensors, and creating the batched Fourier transform for the plane wave computation. The steps are like the ones required to create the previous example. There are some differences between the code snippet in Figure~\ref{alg:fft_classic} and the code snippet in Figure~\ref{alg:fft_spheres}. First, at line $2$ we are required to define the offset array (we do not write the entire code and leave it as $\ldots$ due to space constraints). As outlined in Figure~\ref{fig:spheres}, we project the spheres on the $xy$-plane and use the projection as an offset array. This offset array is like a Compressed Sparse Row (CSR) format because only the $z$ dimension is compressed, while the $x$ and $y$ dimensions are kept as dense. The offset array is used at line $18$ to create the bounding domain for the input tensor. At line $19$ we distribute the $x$ dimension similar to the previous case. Given that, the $z$ dimension has varying lengths, we augment the notation (extra information in the string argument) to allow for dimensions to be merged and even sorted based on the varying length in the $z$-dimension. These details will be described in more detail in the final release of the software.

The second difference is the presence of the batch dimension. Recall that there are multiple wavefunction spheres stacked together. At line $9$-$10$, we specify the lower and upper points for the batch dimension. These points are used at line $17$ and $26$ to specify another dimension in both the input and output domains. The two domains \texttt{dom\_in} and \texttt{dom\_out} are specified as an array of two domains. This corresponds to a larger domain obtained as a cross product between the composing domains. Both tensors \texttt{ti} and \texttt{to} are four-dimensional tensors. Note the order in which the domains are added to \texttt{dom\_in} and \texttt{dom\_out}. The order in which the dimensions are added matters. In this example, we specified the batch dimension as the first dimension, which corresponds to the fastest dimension in memory. If the batch dimension was added after the three-dimensional domains, then the batch dimension would have corresponded to the slowest dimension in memory. More details on the domain creation is in the documentation of the software release.

\section{Experiments and Results}
In the following section, we present experimental results obtained on the HP Cray EX supercomputer. First, we provide details about the configuration flags used to compile and execute the code. Then, we showcase scaling results for two distributed 3D Fourier transforms, outlining the need for flexible implementations.

\subsection{Methodology}

We use the Perlmutter supercomputer at the National Energy Research Scientific Computing Center (NERSC) for all the experiments. Perlmutter provides two partitions users can use, one with CPU-only nodes and one with GPU-accelerated nodes. In this work, we focus on the GPU implementation and only report results on the GPU-accelerated nodes. For this partition, Perlmutter has $1792$ GPU-accelerated nodes, each equipped with an AMD EPYC $7763$ (Milan) CPU and four NVIDIA A$100$ GPUs. Furthermore, Perlmutter features the HPE Slingshot 11 interconnect fabric, which uses a 3-hop dragonfly topology for efficient data transfer. For the strong scaling results, we use up to $256$ physical node and $1024$ GPUs ($4$ GPUs per node). 

We compiled our framework using the NVIDIA toolkit 23.9. We use cuFFT for the local computation. Furthermore, we use CUDA to implement some of the small codelets that pack and rotate the data locally on the GPU before communicating it over the network. The goal is to place the data on the GPU and minimize copies with the host CPU. As such, we use CUDA-aware MPI routines for performing the \texttt{alltoall} communication. This functionality is provided on Perlmutter. In the scenario that this functionality is not available, we also provide support to move data to the host and then initiate a regular MPI \texttt{alltoall} call.

\subsection{Experimental Results}

\begin{figure}[t]
    \centering
    \includegraphics[width=\textwidth]{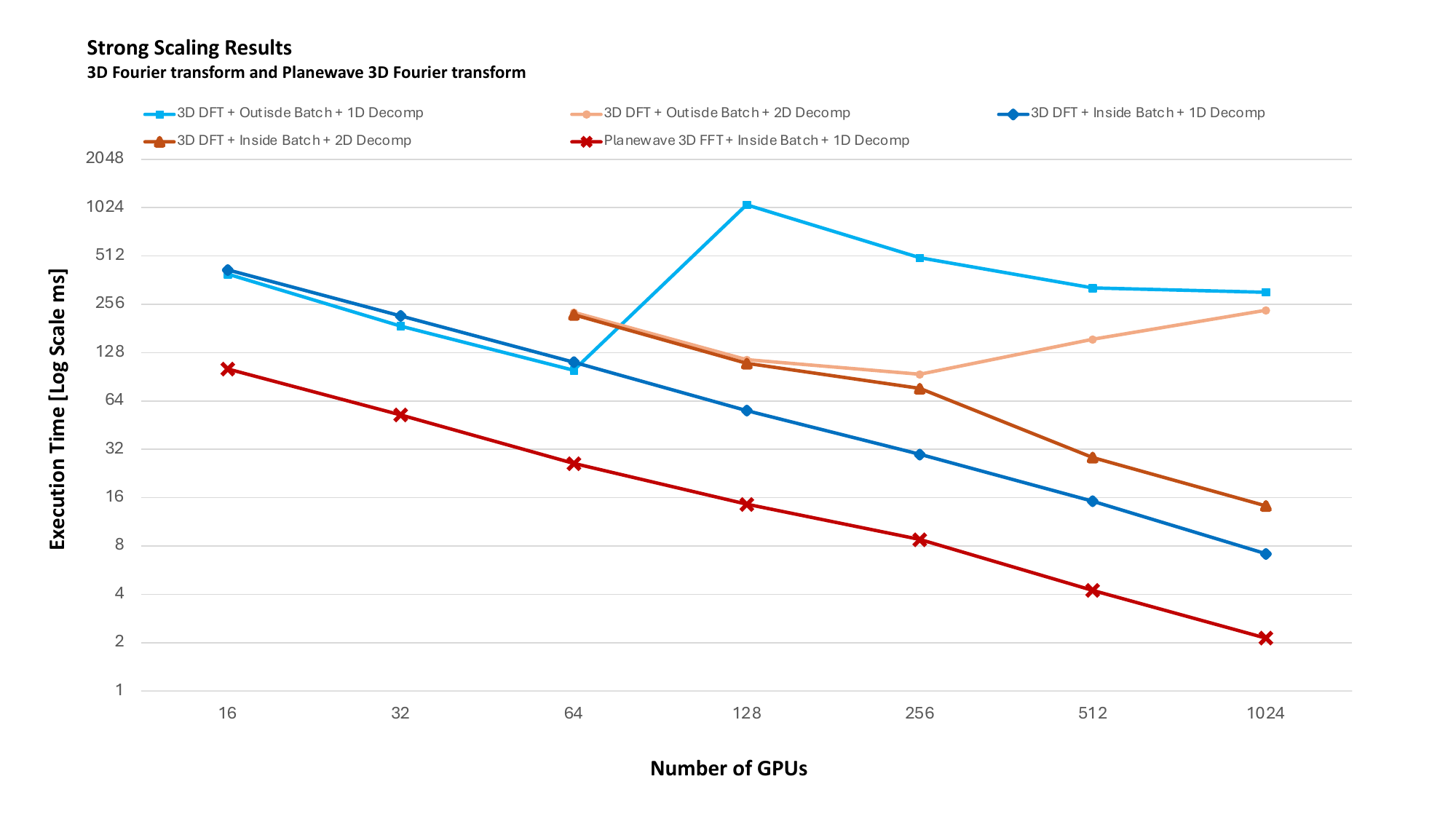}
    \caption{Strong scaling results for different 3D Fourier transform variants implemented within our framework. The red line corresponds to the plane wave transform applied on spheres of diameter $128$. The other four lines correspond to the full 3D Fourier transform applied on a cube of width of $256$ using 1D or 2D processing grids and batching or not the computation.}
    \label{fig:strong_scaling}
    \vspace{-2mm}
\end{figure}

We perform strong scaling experiments, where we keep the problem size fixed and increase the number of parallel processors. All experiments are run multiple times. We execute a warmup phase of $10$ iterations, where we do not measure the execution time. We then execute a hot phase of another $10$ iterations, where we measure the execution time. For all experiments, we take the average of the $10$ iterations in the hot phase. Figure~\ref{fig:strong_scaling} shows strong scaling results for a full 3D Fourier transform and a plane wave 3D Fourier transform. We pick a typical size of $256^3$ for the Fourier transform and $256$ for the batches. The plane wave Fourier transform assumes that the wavefunction sphere has a diameter of $128$. 

The full 3D Fourier transform has four variants. The first variant (dark blue line) uses a 1D processing grid and batches the computation. The second variant (light blue line) uses a 1D processing grid and does not batch the Fourier transforms. Basically, this version loops $256$ times around a distributed 3D Fourier transform. The third variant (dark orange line) uses a 2D processing grid and batches the computation. Finally, the last variant (light orange line) uses a 2D processing without batching. The plane wave 3D Fourier transform uses a 1D processing grid, batches the wavefunctions and gradually pads the data with zeros, keeping communication to a minimum. We first parallelize the data in the dimensions of the Fourier transforms. If the number of processors is greater than the dimensions, we then parallelize in the batch dimension.

There are two important aspects that need to be outlined. First, batching the computation and data movement is important. Both 3D Fourier transforms using 1D and 2D processing grids with no batching experience performance degradation as the number of GPUs is increased. For Fourier transform size of $256^3$, the amount of data that is being transferred is quite small as the number of processors is increased. The light blue line suffers a big jump in execution time when increasing the number of processing units from $64$ to $128$ most likely because the MPI library changes the \texttt{alltoall} algorithm given the size of the data. Batching the computation and data movement alleviates this problem as outlined by the dark blue and orange lines. Second, the planewave Fourier transform scales almost linear to $1024$ GPUs and provides faster execution time compared to the regular 3D Fourier transform that batches (dark blue line). The planewave Fourier transform pads the data in stages, keeping the amount of data moved across the network to a minimum.

\section{Conclusion}
In this work, we have presented FFTB, a flexible and extensible framework for distributed Fourier transforms targeting both CPU and GPU supercomputers. We provided a description of the framework, outlining key characteristics. We constructed FFTB as a modular framework that allows different types of Fourier transforms to be described. In the paper, we focused on classic 3D Fourier transforms applied on data mapped to cuboid grids. Furthermore, we showed that with minor changes, we can extend the framework to plane wave 3D Fourier transforms, needed by a multitude of plane wave Density Functional Theory codes. We also showed multiple implementations with different behaviors on multi-node supercomputers such as Perlmutter, outlining the need for flexible frameworks.  Moving forward, we will focus on integrating our approach with current Density Functional Theory codes. In addition, we will extend the functionality to target other supercomputers like Frontier (AMD GPUs), Aurora (Intel GPUs) and Fugaku (ARM CPUs). We will bundle the implementation and all the extensions making it available on Github.

\section*{Acknowledgements}

This research was supported by the U.S. Department of Energy, Office
of Science, Office of Advanced Scientific Computing Research, Sci-
entific Discovery through Advanced Computing (SciDAC) program
through the FASTMath Institute under Contract No. DE-AC02-
05CH11231 at Lawrence Berkeley National Laboratory. This re-
search used resources of the National Energy Research Scientific
Computing Center (NERSC), a U.S. Department of Energy Office
of Science User Facility operated under Contract No. DE-AC02-
05CH11231. 
\bibliographystyle{abbrv}
\bibliography{biblio}
\end{document}